\documentclass[showpacs,aps,graphicx,twocolumn]{revtex4}
\usepackage{graphicx}

\begin{document}

\title{Efficient quantum cryptography  network without entanglement and quantum memory\footnote{published in
\emph{Chinese Physics Letters} \textbf{23} (11),
 2896-2899 (2006).}}
\author{ Chun-Yan Li$^{1,2}$, Xi-Han Li$^{1,2}$, Fu-Guo Deng$^{1,2,3}$\footnote{E-mail addresses: fgdeng@bnu.edu.cn}, Ping Zhou$^{1,2}$, Yu-Jie Liang$^{1,2}$, Hong-Yu Zhou$^{1,2,3}$}
\address{$^1$ The Key Laboratory of Beam Technology and Material
Modification of Ministry of Education, Beijing Normal University,
Beijing 100875\\
$^2$ Institute of Low Energy Nuclear Physics, and Department of
Material Science and Engineering, Beijing Normal University,
Beijing 100875\\
$^3$ Beijing Radiation Center, Beijing 100875}
\date{\today }

\begin{abstract}
An efficient quantum cryptography network protocol is proposed with
$d$-dimension polarized photons, without resorting to entanglement
and quantum memory. A server on the network, say Alice, provides the
service for preparing and measuring single photons whose initial
state are $\vert 0\rangle$. The users code the information on the
single photons with some unitary operations. For preventing the
untrustworthy server Alice from eavesdropping the quantum lines, a
nonorthogonal-coding technique (decoy-photon technique) is used in
the process that the quantum signal is transmitted between the
users. This protocol does not require the servers and the users to
store the quantum state and almost all of the single photons can be
used for carrying the information, which makes it more convenient
for application than others with present technology. We also discuss
the case with a faint laser pulse.
\end{abstract}
 \maketitle

Preventing a vicious eavesdropper, say Eve from stealing the message
in communication is one of the most important issues nowadays. In
the classical communication, the security of the public key
crypto-systems is generally based on their computational complexity.
For example, the security of the Rivest-Shamir-Adleman public key
scheme \cite{RSA} depends on the difficulty of factoring a large
integer. Up to now, none of them has been proven to be
unconditionally secure. The Vernam one-time pad crypto-system
\cite{Vernam} provides a secure way for two remote parties to
communicate with a private key which is required to be long as the
message and can only be used one time securely. As a classical
signal is in one of the eigenvectors of a operator, it can be copied
fully and freely. Quantum cryptography or quantum key distribution
(QKD) \cite{Book,Gisin} provides a secure way for creating a private
key between two authorized users, and becomes one of the most
important applications of quantum information \cite{Book,Gisin}. For
instance, Bennett and Brassard \cite{BB84} presented an original QKD
protocol, called BB84, with four nonorthogonal single-photon states
in 1984 , and Ekert \cite{Ekert91} introduced a QKD protocol based
on the correlation of a maximally entangled two-particle quantum
system, an Einstein-Podolsky-Rosen(EPR) pair in 1991. Now, there is
much attention focused on QKD
\cite{Gisin,BB84,Ekert91,BBM92,B92,ABC,LongLiu,CORE,BidQKD}. It has
been also well developed in experimental implementations
\cite{Gisin}.

In recent years, the any-to-any QKD protocols for the secure
communication on a passive optical network, which is a requirement
of practical implementations, have been studied by some groups
\cite{Phoenix,Townsend,Biham,MUQKDguo,DLMXL,LZWD,LIXHnetwork}.
Phoenix \emph{et al}. \cite{Phoenix} proposed a multi-user QKD
scheme with single photons in 1995 following the ideas in Bennett
1992 protocol \cite{B92} and BB84 QKD protocol \cite{BB84}. In this
scheme, half of the quantum information carriers (QIC) are useful
for carrying the information if they remove the ideas from BB84 QKD.
Its efficiency for qubits $\eta_q\equiv \frac{q_u}{q_t}=50\%$, the
same as that in BB84 QKD. Here $q_u$ is useful qubits and $q_t$ is
total qubits used. In 1997 Townsend \cite{Townsend} demonstrated the
multi-user QKD (MUQKD) on an optical fiber networks with faint laser
pulses following the ideas in BB84 QKD \cite{BB84}. Biham \emph{et
al}. \cite{Biham} proposed a MUQKD protocol with quantum memories in
1996. In their MUQKD scheme, no more than $\frac{1}{8}$ QIC can be
used as the qubits for the raw key. The advantage is that the users
on the network can work without quantum channels if they store the
QIC in the quantum memories in advance \cite{Biham}. Xue \emph{et
al} \cite{MUQKDguo} presented a way for MUQKD using the mixture of
single photons and EPR pairs as the QIC. The efficiency $\eta_q$ was
improved to approach 100\% with the ideas in Ref. \cite{ABC}.
Another two MUQKD schemes \cite{DLMXL,LZWD} were presented by
modifying the quantum dense coding \cite{quantumdensecoding} and the
point-to-point QKD protocol proposed by Long and Liu \cite{LongLiu}.
In these two MUQKD schemes, the QIC are EPR pairs. Moreover, their
efficiency $\eta_q$ is improved to approach 100\% only when the
users or the servers on the network exploit quantum memory to store
the QIC. Although the technique of quantum storage is a vital
ingredient for quantum information and there has been a great deal
of interests in developing it \cite{storage}, it cannot be used in
the practical application at present.

In this Letter, we will introduce a new multi-user QKD network
protocol without resorting to entanglement and quantum memory. The
$d$-dimensional single photons are prepared and measured by the
server Alice on the network with one measuring basis (MB). The users
code the information on the single photons with some unitary
operations, and each photon can carry $log_2 d$ bits of information.
Almost all the photons can be used to carry the useful information,
the efficiency for qubts $\eta_q$ approaches 100\%. The users can
exploit some decoy photons (in nonorthogonal states) which are
obtained by operating some samples with a Hadamard operation to
ensure the security of the quantum communication. We also discuss
this multi-user network with a faint laser pulse.

\begin{figure}[!h]
\begin{center}
\includegraphics[width=8cm,angle=0]{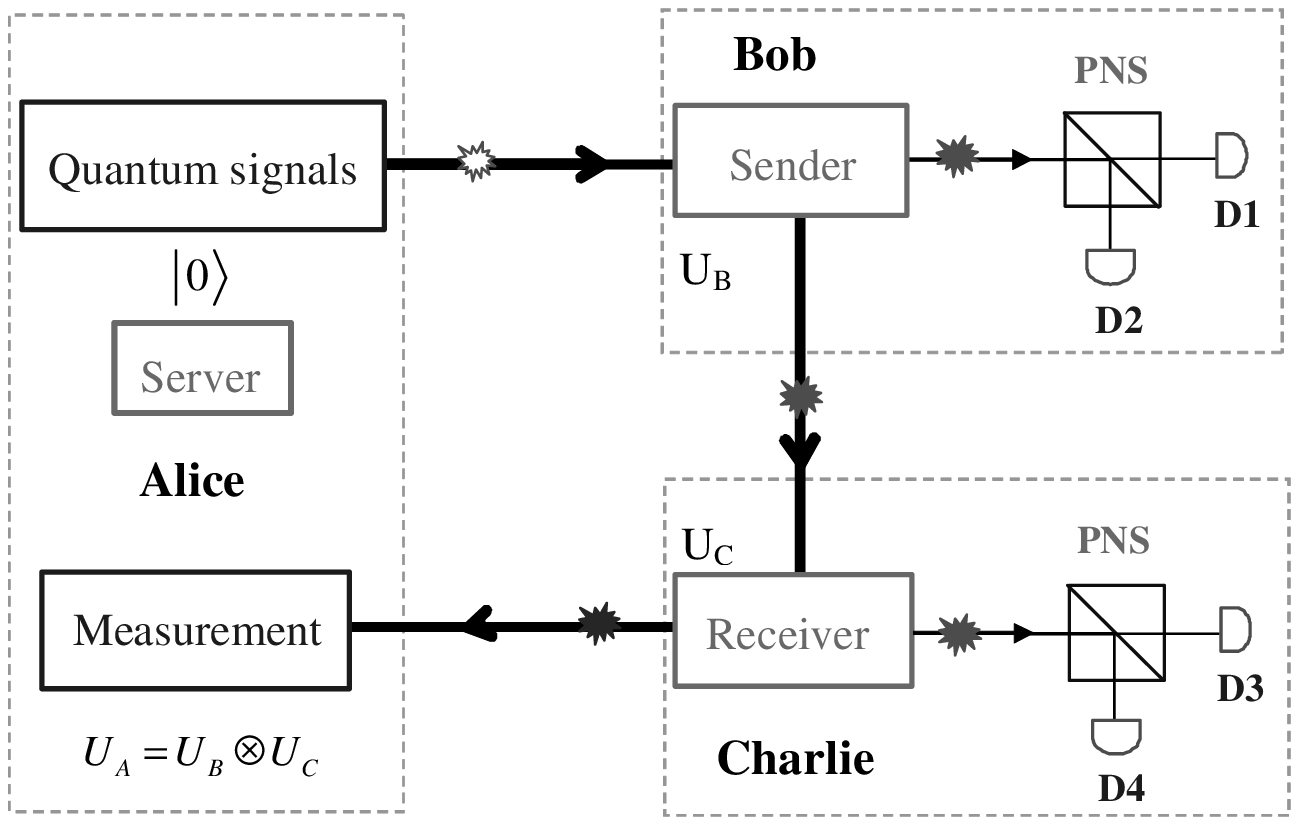} \label{f1}
\caption{ The subsystem of the network in this MUQKD scheme, similar
to those in Refs.
\cite{Phoenix,Townsend,Biham,MUQKDguo,DLMXL,LZWD,LIXHnetwork}. PNS:
photon number splitter; $D_m$ ($m=1,2,3,4$) are four single-photon
detectors. $U_B$ and $U_C$ are the operations done by Bob and
Charlie, respectively.}
\end{center}
\end{figure}

We use the same structure of the network as those in Refs.
\cite{Phoenix,Townsend,Biham,MUQKDguo,DLMXL,LZWD} in the present
MUQKD network protocol, i.e., its subsystem (a cell of the QKD
network) can be simplified to three parts, the server (Alice), the
sender (Bob) and the receiver (Charlie). All the cells build up a
practical network. A MUQKD scheme is explicit if the principle of
its subsystem is described clearly
\cite{Phoenix,Townsend,Biham,MUQKDguo,DLMXL,LZWD,LIXHnetwork}.

A subsystem in our MUQKD scheme is shown in Fig.1.  Alice provides
the service for preparing and measuring the polarized
$d$-dimensional single photon $T$. For a $d$-dimensional single
photon, we can choose two nonorthogonal MBs as
 $Z_{d}$ and $X_d$ \cite{dengepl}. The MB $Z_{d}$ which has $d$
eigenvectors can be written as:
\begin{eqnarray}
\left\vert  0 \right\rangle, \;\;\;\left\vert  1 \right\rangle,
\;\;\;\;\left\vert  2 \right\rangle, \;\; \cdots, \;\;\;\;\left\vert
{d - 1} \right\rangle.
\end{eqnarray}
The $d$ eigenvectors of the MB $X_{d}$ can be described as
\begin{eqnarray}
\vert 0\rangle_x&=&\frac{1}{{\sqrt d }}\left( {\left\vert  0
\right\rangle + \vert 1\rangle \;\; + \cdots \;\; + \left\vert
{d-1}\right\rangle }\right),\;\nonumber \\
\vert 1\rangle_x&=&\frac{1}{{\sqrt d }}\left({\left\vert  0
\right\rangle + e^{{\textstyle{{2\pi i} \over d}}} \left\vert  1
\right\rangle + \cdots
+ e^{{\textstyle{{(d-1)2\pi i} \over d}}} \left\vert  {d-1} \right\rangle} \right),\; \nonumber\\
\vert 2\rangle_x&=&\frac{1}{{\sqrt d }}\left({\left\vert  0
\right\rangle + e ^{{\textstyle{{4\pi i} \over d}}} \left\vert 1
\right\rangle + \cdots + e^{{\textstyle{{(d-1)4\pi i} \over d}}}
\left\vert  {d-1} \right\rangle }\right),\nonumber\\
&&\cdots \cdots \cdots \cdots \cdots \cdots \nonumber\\
\vert d-1\rangle_x&=&\frac{1}{{\sqrt d }}(\left\vert  0
\right\rangle + e ^{{\textstyle{{2(d-1)\pi i} \over d}}} \left\vert
1 \right\rangle  + e ^{{\textstyle{{2\times 2(d-1)\pi i} \over d}}}
\left\vert 2 \right\rangle + \cdots \nonumber\\
&& + e^{{\textstyle{{(d-1)\times 2(d-1)\pi i} \over d}}} \left\vert
{d-1} \right\rangle ).
\end{eqnarray}
The two vectors $\vert k\rangle$ and $\vert l\rangle_x$ coming from
two MBs satisfy the relation $\vert \langle k|l\rangle_x \vert
^2=\frac{1}{d}$. We can use the unitary operation $U_j$
($j=0,1,\cdots, d-1$) to transfer the state $\vert 0\rangle$ into
another state $\vert j\rangle$, i.e., $U_j \vert 0\rangle = \vert
j\rangle$.
\begin{equation}
U_{j} = \vert j \rangle \langle 0\vert.
\end{equation}
Moreover, the $d$-dimensional Hadamard ($H_d$) operation can
transfer an eigenvector of the MB $Z_d$ into that of the MB $X_d$,
i.e., $H_d\vert j\rangle=\vert j\rangle_x$. Here \cite{dengepl}
\begin{eqnarray}
H_d  =\frac{1}{\sqrt{d}} \left( {\begin{array}{*{20}c}
   1 & 1 &  \cdots  & 1  \\
   1 & {e^{2\pi i/d} } &  \cdots  & {e^{(d-1)2\pi i/d} }  \\
   1 & {e^{4\pi i/d} } &  \cdots  & {e^{(d-1)4\pi i/d}  }\\
    \vdots  &  \vdots  &  \cdots  & \vdots  \\
   1 & {e^{2(d-1)\pi i/d} } &  \cdots  & {e^{(d-1)2(d-1)\pi i/d} }  \\
\end{array}} \right)\label{HD}.
\end{eqnarray}

For simplifying the process of error rate analysis, the traveling
single photon $T$ is prepared by the server Alice initially in the
state $\vert 0\rangle_z=\vert 0\rangle$ in each round. That is, all
the users including the server Alice agree that the original state
of the traveling single photon $T$ is $\vert 0\rangle$. Alice sends
the photon $T$ to the sender Bob. Bob chooses two modes, the
checking-eavesdropping mode and the message-coding mode, for the
photon received with the probabilities $1-P_{bm}$ and $P_{bm}$,
respectively, similar to Refs. \cite{twostep}. When he chooses the
checking-eavesdropping mode, Bob measures the photon with the MB
$Z_d$. When he chooses the message-coding mode, Bob codes the photon
$T$ by choosing  randomly one of the $d$ unitary operations
$\{U_j\}$, say $U_B$. Moreover, Bob should exploit a
nonorthogonal-coding technique (i.e., decoy-photon technique) to
determine whether an eavesdropper is monitoring the quantum line
between the two users. That is, Bob should replace the photon $T$
with a decoy one in a nonorthogonal state by using a probability
$P_d$ ($< \frac{1}{2}$) before he sends it to the receiver Charlie.
In detail, he can prepare the decoy photon by performing a $H_d$
operation on the traveling photon $T$ after coding it with one of
the  unitary operations $\{U_j\}$ randomly (The
 decoy photon in this scheme is different from the
decoy state in Ref. \cite{wangxb}. It is just a photon in a
nonorthogonal state, compared with its original state, not the faint
pulses with different intensities.). In this way, the decoy photon
is randomly in one of the $d$ states $\{\vert 0\rangle_x, \vert
1\rangle_x, \cdots, \vert d-1\rangle_x\}$. After receiving the
photon $T$, Charlie operates it similar to Bob. That is, Charlie
performs the operation $U_C \in \{U_{mn}=\vert m\rangle \langle
n\vert; \; m, n=0,1,\cdots, d-1 \}$ on the photon and then sends it
to the server Alice if Charlie chooses the message-coding mode,
otherwise he measures the photon with one of the two MBs $Z_d$ and
$X_d$ by using the probabilities $P_{cz}$ and $P_{cx}$,
respectively. If Alice received the photon $T$, she measures it with
the MB $Z_d$ and publishes the difference between its original state
and the final one. After Bob deletes the results coming from the
decoy photons measured by Alice, Charlie can obtain the outcomes
$U_A=U_B \otimes U_C$.

For preventing the eavesdropper from stealing the information about
the operations $U_B$ done by Bob with a multi-photon signal
\cite{multiphotonattack}, Bob should check the number of the photons
in each signal. That is, Bob should analyse the probability that the
case in which there are more than one photon in the signal takes
place. This task can be completed by sampling a subset of signals
randomly and measuring them with two single-photon detectors after
splitting them with a photon number splitter (PNS), see Fig.1. In
fact, the check done by Bob is just used to determine whether the
untrustworthy server Alice inserts a Trojan horse in the original
signal. Certainly,  Bob should use a special filter (just the
photons with the special frequency can penetrate it \cite{Gisin}) to
filtrate the light from background or a fake signal \cite{dengcpl}
before he operates the photons. The receiver Charlie should also do
the operation same as Bob to prevent  Alice from eavesdropping with
a Trojan horse attack.

With the decoy photons and PNSs, Bob and Charlie can check the
security of their quantum communication by analyzing a large enough
subset of the results. As the initial state of the photons is $\vert
0\rangle$, the analysis of the error rate done by Bob an Charlie
does not need the help of the server Alice. Bob and Charlie can
check eavesdropping efficiently with a refined error analysis
technique same as that in Ref. \cite{ABC}. Thus this MUQKD protocol
can be made to be secure.

Now let us discuss several issues. Firstly, the requirement that the
travelling photon $T$ is initially in the state $\vert 0\rangle$ is
useful for improving the security of this MUQKD protocol against
dishonest servers. If the photon $T$ is randomly in one of the
states $\{\vert j\rangle\}$, the error rate analysis of the samples
transmitted from  Bob to Charlie needs the help of the server Alice.
In this way, Alice can eavesdrop the operations done by Bob and
Charlie fully and freely, and hide her attack with a cheat. We use
the case with a two-dimensional polarized single photon to describe
the principle of this attack. In detail, we assume that the state of
the photon $T$ is $\vert \psi'\rangle_{T} \in \{\vert 0\rangle,
\vert 1\vert, \vert +x\rangle=\frac{1}{\sqrt{2}}(\vert 0\rangle +
\vert 1\rangle), \vert -x\rangle=\frac{1}{\sqrt{2}}(\vert 0\rangle
-\vert 1\rangle)\}$. Alice intercepts the photon $T$ after it is
operated by the sender Bob, and stores it. She sends one photon in
an EPR pair in the state $\vert
\psi^-\rangle_{AB}=\frac{1}{\sqrt{2}}(\vert 01\rangle -\vert
10\rangle)_{AB}$ to Charlie, say the photon $B$, instead of the
original one $T$. If Charlie chooses the message-coding mode on the
photon $B$, Alice measures the photon $T$ with the MB $Z$ and
performs a Bell-basis measurement on the EPR pair. Obviously, she
can obtain all the information about the operations $U_B$ and $U_C$
because Charlie only chooses one of the two operations $U_0=\vert
0\rangle\langle 0\vert + \vert 1\rangle\langle 1\vert$ and
$U_1=\vert 0\rangle\langle 1\vert + \vert 1\rangle\langle 0\vert$
which make the EPR pair in the states $\vert \psi^-\rangle_{AB}$ and
$\vert \phi^-\rangle=\frac{1}{\sqrt{2}}(\vert 00\rangle - \vert
11\rangle)_{AB}$, respectively. If Charlie chooses the
checking-eavesdropping mode, Alice performs a Bell-basis measurement
on the photons $A$ and $T$. It is well known that the state of the
photon $B$ measured by Charlie is correlated to the results of the
Bell-basis measurements \cite{teleportation}. That is, if the
results are $\vert \psi^-\rangle_{AT}$, $\vert \psi^+\rangle_{AT}$,
$\vert \phi^-\rangle_{AT}$ and $\vert \phi^+\rangle_{AT}$, Alice
needs only publish a fake information about the initial state of the
photon $T$ after the unitary operations $I=U_0$, $\sigma_z=\vert
0\rangle\langle 0\vert - \vert 1\rangle\langle 1\vert$,
$\sigma_x=U_1$ and $i\sigma_y=\vert 0\rangle\langle 1\vert - \vert
1\rangle\langle 0\vert$, respectively \cite{teleportation}. Here
$\vert \psi^+\rangle_{AT}=\frac{1}{\sqrt{2}}(\vert 01\rangle +\vert
10\rangle)_{AT}$ and $\vert
\phi^+\rangle_{AT}=\frac{1}{\sqrt{2}}(\vert 00\rangle +\vert
11\rangle)_{AT}$. Fortunately, in our MUQKD protocol, the users can
accomplish the error rate analysis without the help of the server,
which makes the attack invalid.

Secondly, different from Ref. \cite{ABC}, Charlie can choose the MB
$X_d$ with a large probability when he chooses the
checking-eavesdropping mode for obtaining more correlated outcomes.
For the symmetry, we assume that the outcome useful obtained with
the MB $Z_d$ is equal to that with the MB $X_d$, i.e.,
\begin{eqnarray}
(1-p_d)P_{bm}P_{cm}P_{cz}=P_{d}P_{bm}P_{cm}P_{cx},
\end{eqnarray}
where $P_{bm}$ and $P_{cm}$ are the probabilities that Bob and
Charlie choose the message-coding mode, respectively; $P_{cz}$ and
$P_{cx}=1-P_{cz}$ are the probabilities that Charlie measures his
samples with the MB $Z_d$ and $X_d$, respectively. That is, when
$P_{cz}=P_{d}$, the probability that Bob and Charlie obtain the
correlated outcomes of the samples approaches the maximal value
$P_{eu}=2(1-P_d)P_d$.

Thirdly, let us discuss the case that our MUQKD protocol is
implemented with a practical faint laser pulse. The probability that
there are $n$ photons in a pulse follows the Poisson statistics
\cite{Gisin},
\begin{equation}
P(n, \mu)=\frac{\mu ^{n}}{n!}e^{-\mu }  \label{p1}
\end{equation}%
where $n$ is the number of photons in a coherent state and $\mu
=\left\langle n\right\rangle $ is the mean photon number. Then the
probabilities that a non-empty weak coherent pulse contains more
than one photon is \cite{Gisin}
\begin{eqnarray}
P(n >1|n>0, \mu)&=&\frac{1-P(0, \mu)-P(1, \mu)}{1-P(0, \mu)}  \nonumber \\
&=&\frac{1-(1+\mu )e^{-\mu }}{1-e^{-\mu }}\cong \frac{\mu }{2}.
\label{p2}
\end{eqnarray}
If $\mu =0.05$, the probabilities $P(n>1|n>0, \mu=0.05)\approx
2.5\%$. That is, when the Fock states are attenuated to one photon
per 20 pulses, the probability that there are more than one photon
in a pulse is about $2.5\%$.

The instances with more than one photons in a pulse will decrease
the security of this MUQKD protocol. The reason is that the
dishonest server can split one photon from the multi-photon pulse
operated by Bob and measure it with the MB $Z_d$. In this way, Alice
can get all the useful information about the operation $U_B$. With
the outcome $U_A$, she can obtain the private key fully and freely.
In order to prevent Alice from stealing the information with PNS
attack, the probability $P_{cu}$ that the receiver Charlie obtains
an useful outcome when he measures a sample photon run from Bob is
by far larger than the probability $P(n>1|n>0, \mu=0.05)$, i.e.,
\begin{eqnarray}
P_{cu}=\eta_{opt}\eta_{d} \gg P(n >1|n>0, \mu=0.05)=2.5\%,
\end{eqnarray}
where $\eta_{opt}$ and $\eta_{d}$ are the efficiency of the
transmission on a fibre and that of a detector, respectively.
Otherwise, Alice can steal some of the information about the key
with a better quantum channel. In detail, Alice, on one hand,
intercepts all the single-photon pulse and discards them. On other
hand, she splits the multi-photon signal with some PNSs and sends
one of the photons in the pulse to Charlie with a nearly ideal
channel in which the loss is very low. Obviously, her eavesdropping
does not introduce errors in the outcomes of the samples chosen by
Charlie. Moreover, the loss of the signal is compensated with a good
channel. Thus Bob and Charlie cannot detect Alice's vicious action.
But the story is changed when $P_{cu} \gg P(n >1|n>0, \mu=0.05)$. In
this time, Bob and Charlie can monitor the number of the photons in
each signal by sampling some pulses randomly and measuring them
after splitting with some PNSs. On the other hand, Bob and Charlie
can exploit privacy amplification to distil a short key privately
\cite{Gisin}.

Compared with the MUQKD protocols existing
\cite{Phoenix,Townsend,Biham,MUQKDguo,DLMXL,LZWD,LIXHnetwork}, this
one requires the users on the network to have the capability of
measuring single photons and unitary operations, not Bell-basis
measurement and quantum memory, which makes it more convenient in
application. Moreover, the efficiency for qubits $\eta_q$ approaches
100\% as almost all the photons can be used to generating the
private key but those for checking eavesdropping (its number is
negligible). The total efficiency $\eta_t\equiv \frac{q_u}{q_t+
b_t}$ also approaches the maximal value 50\% as Alice need only
publish one bit of classical information for each useful qubit,
i.e., $q_u=q_t=b_t=1$. Although the technique for splitting some
photons is in developing \cite{Gisin}, the users can use photon beam
splitter (PBS) to replace PNS for determining the probability that
there are more than one photon in each signal.

In summary, we have presented a MUQKD network protocol without
entanglement and quantum memory. The users on the network exploit
some unitary operations to code their information on a travelling
photon. As the initial state of the photon prepared by the server is
$\vert 0\rangle$, the sender can perform a Hadamard operation on the
photon operated to produce a decoy one which is used to forbid the
dishonest server to eavesdrop freely. With some PNSs, this MUQKD
network protocol can be made to be secure. The efficiency for qubits
and the total efficiency both approach the maximal values, and then
this protocol is an optimal one. Moreover, we discuss the case with
a faint laser pules.

This work was supported by the National Natural Science Foundation
of China under Grant Nos. 10447106, 10435020, 10254002 and A0325401,
and Beijing Education Committee under Grant No. XK100270454.

\end{document}